\documentclass[aps,pre,twocolumn,superscriptaddress,showpacs]{revtex4-2}

\usepackage{textcomp}
\usepackage{amsmath,amstext,amsfonts,amssymb}
\usepackage{graphicx}
\usepackage{gensymb}
\usepackage{float}
\usepackage{xcolor}
\usepackage{lineno}

\newcommand{\bra}{\langle}
\newcommand{\ket}{\rangle}

\newcommand{\Rev}[1]{\textcolor{black}{#1}}

\begin{document}

\title{Properties of carbon up to 10 million kelvin from Kohn-Sham density functional theory molecular dynamics}

\author{Mandy Bethkenhagen}
\email[]{mandy.bethkenhagen@ens-lyon.fr}
\affiliation{Lawrence Livermore National Laboratory, Livermore, California 94550, USA}
\affiliation{\'{E}cole Normale Sup\'{e}rieure de Lyon, Universit\'e Lyon 1, Laboratoire de G\'eologie de Lyon, CNRS UMR 5276, 69364 Lyon Cedex 07, France}

\author{Abhiraj Sharma}
\affiliation{College of Engineering, Georgia Institute of Technology, Atlanta, Georgia 30332, USA}

\author{Phanish Suryanarayana}
\affiliation{College of Engineering, Georgia Institute of Technology, Atlanta, Georgia 30332, USA}

\author{John E. Pask}
\affiliation{Lawrence Livermore National Laboratory, Livermore, California 94550, USA}

\author{Babak Sadigh}
\affiliation{Lawrence Livermore National Laboratory, Livermore, California 94550, USA}

\author{Sebastien Hamel}
\affiliation{Lawrence Livermore National Laboratory, Livermore, California 94550, USA}

\date{\today}

\begin{abstract}
Accurately modeling dense plasmas over wide ranging conditions of pressure and temperature is a grand challenge critically important to our understanding of stellar and planetary physics as well
as inertial confinement fusion. In this work, we employ Kohn-Sham density functional theory (DFT) molecular dynamics (MD) to compute the properties of carbon at warm and hot dense matter conditions in the vicinity of the principal Hugoniot.  In particular, we calculate the equation of state (EOS), Hugoniot, pair distribution functions, and diffusion coefficients for carbon at densities spanning 8~g/cm$^3$ to 16~g/cm$^3$ and temperatures ranging from 100~kK to 10~MK \Rev{using the Spectral Quadrature (SQ) method}. We find that the computed EOS and Hugoniot are in good agreement with path integral Monte Carlo results and the SESAME database. Additionally, we calculate the ion-ion structure factor and viscosity for selected points. All results presented are at the level of full Kohn-Sham DFT MD, free of empirical parameters, average-atom, and orbital-free approximations employed previously at such conditions. 
\end{abstract}




\maketitle


 \section{Introduction}
%

Carbon is one of the most abundant heavy elements found in the interior of stars~\cite{Chabrier1997} and therefore one of the most studied elements in warm dense matter physics~\cite{Gaffney2018, Grabowski2020, Bonitz2020}. Of particular interest are the equation of state (EOS) and transport properties, as they are required for accurately modeling stars and experiments targeting the recreation of stellar interiors in the laboratory. 

For many decades, modeling relied on analytical free-energy models with material properties derived from the theory of partially ionized plasmas~\cite{Fontaine1977, Potekhin2005}. A significant improvement has been made with the development of Kohn-Sham density functional theory (DFT) \cite{Hohenberg1964, Kohn1965}, which has enabled robust and predictive calculations of a wide range of materials properties from the first principles of quantum mechanics, with no empirical parameters.  It has been extensively applied to study the carbon phase diagram up to about 100~kK~\cite{Correa2006, Benedict2014, Vorberger2020} covering the range available to experiments up to the Gbar range~\cite{Nagao2006, Bradley2004, Bradley2009, Gregor2017, Millot2018}. While it is possible to push  standard orbital-based DFT methods beyond this limit for high-density plasmas, as has been demonstrated for hydrogen~\cite{Recoules2009} and recently for carbon~\cite{Bethkenhagen2020}, the comparably low-density range below a compression ratio of 5 remains inaccessible. In particular, the $\mathcal{O}(N^3)$ bottleneck with respect to the number of atoms/electronic states makes standard orbital-based Kohn-Sham DFT particularly expensive at high temperature, especially at low densities, even for small to moderate-sized systems~\cite{Blanchet2020}. Therefore, such conditions have remained generally inaccessible at this level of theory.

To overcome this practical limitation, approaches such as path integral Monte Carlo (PIMC)~\cite{Driver2012, Driver2017},
orbital-free molecular dynamics (OFMD)~\cite{Lambert2006}, pseudoatom molecular dynamics (PAMD)~\cite{Starrett2015, Starrett2016}, and extended first-principles molecular dynamics~\cite{Zhang2016} have been proposed. 
However, each of these methods cannot treat lower temperature conditions on the same level of theory leading to the use of a patchwork of electronic structure methods to describe different conditions of pressure and temperature for a number of materials including carbon~\cite{Benedict2014, Danel2018, Gaffney2018, Swift2020, Militzer2021}, the system of interest in this work. This in turn requires careful stitching of equation of state data and the use of {\em ad hoc} switching functions in the regions where different methods overlap. A consistent ab initio description that overcomes this problem and can be used to establish/benchmark the region of applicability and accuracy of traditional models such as the one-component plasma and the Yukawa model~\cite{Clerouin2016}, or numerical approaches such as the hypernetted chain approximation~\cite{DharmaWardana2008, Bredow2013}, is therefore desirable. Moreover, such ab initio data can be used to inform models~\cite{Bredow2015} and train machine-learned force fields, as recently done for carbon at extreme conditions \cite{willman2022machine}.

The recently developed $\mathcal{O}(N)$ Spectral Quadrature (SQ) method \cite{Suryanarayana2013, Pratapa2016} for large-scale Kohn-Sham DFT calculations, as implemented in the SQDFT code \cite{Pratapa2016,Suryanarayana2018,Sharma2020}, \Rev{overcomes the bottlenecks with respect to both temperature and system size of diagonalization-based $\mathcal{O}(N^3)$ Kohn-Sham methods~\cite{Blanchet2020}}, allowing for a comprehensive and seamless quantum mechanical investigation over the full range of temperatures and systems sizes required. The SQ method formulates DFT densities, energies, forces, and stresses as spectral integrals, yielding a linearly scaling method whose cost decreases with increasing temperature as the Fermi-Dirac distribution becomes smoother and the density matrix becomes more localized. This allows us to study systems at ultra-high temperatures in addition to making the method linear-scaling with the number of atoms. The SQ method extends the use of many-particle Kohn-Sham DFT-MD electronic structure calculations to temperatures and pressures where results can be directly compared with high temperature methods such as PIMC\Rev{~\cite{Zhang2019}. Its application facilitates the construction of wide-range EOS~\cite{Zhang2019, Wu2021}, parametrization of effective one-component plasma models~\cite{Clerouin2022}, and enables the calculation of transport properties in the ultra-high temperature regime~\cite{Sharma2020}}.

Leveraging this methodology, we calculate the Hugoniot for carbon entirely based on Kohn-Sham DFT-MD from the condensed matter regime up to the warm and hot dense matter regime up to 10~MK. Additionally, we present benchmark calculations for selected EOS points in comparison to the widely-used planewave Kohn-Sham DFT codes VASP~\cite{Kresse1993a,Kresse1994,Kresse1996} and PWscf~\cite{Giannozzi2009,Giannozzi2017}, as well as the SESAME 7831 EOS and PIMC data. We also calculate structural and ionic transport properties. In particular, we investigate the pair distribution function and the structure factor at an isochore that corresponds to approximately 3-fold compression. The density and temperature dependence of the diffusion coefficient is explored and we calculate the viscosity close to the maximum compression of the Hugoniot. 


\section{Methods}
\subsection{Spectral Quadrature (SQ) method}
The Spectral Quadrature (SQ) method~\cite{Suryanarayana2013} is a density-matrix based $O(N)$ method for the solution of the Kohn-Sham equations that is particularly well suited for calculations at high temperature. In the SQ method, all quantities of interest, such as energies, forces, and stresses, are expressed as bilinear forms or sums of bilinear forms which are then approximated by quadrature rules that remain spatially localized by exploiting the locality of electronic interactions in real space \cite{Prodan2005}, i.e., the exponential decay of the density matrix at finite temperature~\cite{Goedecker1998,Ismail1999,Benzi2013,Suryanarayana2017}. In the absence of truncation, the method becomes mathematically equivalent to the recursion method \cite{Haydock1972,Haydock1975} for the choice of Gauss quadrature, while for Clenshaw-Curtis quadrature, the classical Fermi operator expansion (FOE) \cite{Goedecker1994,Goedecker1995} in Chebyshev polynomials is recovered. Being formulated in terms of the finite-temperature density matrix, the method is applicable to metallic and insulating systems alike, with increasing efficiency at higher temperature as the Fermi operator becomes smoother and density matrix becomes more localized \cite{Pratapa2016}. $O(N)$ scaling is obtained by exploiting the locality of the density matrix at finite temperature, while the exact diagonalization limit is obtained to desired accuracy with increasing quadrature order and localization radius. Convergence to standard $O(N^3)$ planewave results, for metallic and insulating systems alike, is readily obtained \cite{Pratapa2016,Suryanarayana2018}. 

While mathematically equivalent to classical FOE methods in the absence of truncation for a particular choice of quadrature, the more general SQ formulation affords a number of advantages in practice~\cite{Pratapa2016,Suryanarayana2018}. These include: 
(1) The method is expected to be more robust since it explicitly accounts for the effect of truncation on the Chebyshev expansion.
(2) The method computes only the elements of the density matrix needed to evaluate quantities of interest --- e.g., diagonal elements to obtain the electron density, and only those off-diagonal elements that correspond to nonzero values in the nonlocal pseudopotential projectors for the nonlocal atomic forces --- rather than computing 
the full density matrix (to specified threshold) as in FOE methods. 
(3) The method computes the Fermi energy without storage or recomputation of Chebyshev matrices as required in FOE methods. 
(4) The method admits a decomposition of the global Hamiltonian into local sub-Hamiltonians in real space, reducing key computations to local sub-Hamiltonian matrix-vector multiplies rather 
than global full-Hamiltonian matrix-matrix multiplies as in FOE methods. Since the associated local multiplies are small (according to the decay of the density matrix) and independent of one 
another, the method is particularly well suited to massively parallel implementation; whereas the global sparse matrix-matrix multiplies required in FOE methods pose significant challenges 
for parallel implementation \cite{Bowler2012}.

As discussed above, the SQ method circumvents the calculation of the Kohn-Sham orbitals/eigenvalues and directly evaluates the quantities of interest such as energies, forces, and stresses through spatially localized quadrature rules. Consequently, properties that explicitly depend on orbitals and/or eigenvalues, such as thermal and electrical conductivities, and cannot be expressed in terms of analytic functions of the density matrix, are not readily obtainable by the present SQ formulation.

\subsection{Numerical details}
In the present work, we employ the massively parallel SQDFT code \cite{Suryanarayana2018} for high-temperature Kohn-Sham calculations. SQDFT implements the SQ method in real space using a high-order 
finite difference discretization, wherein sub-Hamiltonians are computed and applied for each finite-difference grid point. For efficient molecular dynamics (MD) simulations, Gauss quadrature 
is employed for the calculation of density and energy in each SCF iteration whereas Clenshaw-Curtis quadrature is employed for the calculation of atomic forces and stress tensor~\cite{Pratapa2016,Sharma2020}. 
The SQDFT molecular dynamics simulations were carried out for a series of 64-atom C unit cells at densities between 8 and 16~g/cm$^3$ and temperatures ranging from 100~kK to 10~MK. We employ two different optimized norm conserving Vanderbilt (ONCV) \cite{Hamann2013} pseudopotentials depending on the temperature, i.e., we consider only the 2s2p-valence below 750~kK and switch to an all-electron ONCV pseudopotential for higher temperatures to correctly account for the partial ionization of the 1s states. Exchange and correlation were modeled in the local density approximation (LDA)~\cite{Perdew1981}. NVT simulations were carried out using a Nos\'{e}-Hoover thermostat \cite{Nose1984,Hoover1985} with $\sim$200--2~000 steps for equilibration followed by $\sim$3~000--30~000 steps of production. The timestep of the simulations was chosen between 0.2~fs for low temperatures and 0.02~fs for high temperatures. A finite difference grid spacing of $\sim$0.1~\Rev{bohr} (commensurate with unit cell dimensions), Gauss and Clenshaw-Curtis quadrature orders of 16--80 and 32--120, respectively, and localization radius of 0.5--3.5~\Rev{bohr} were employed in the SQ calculations to obtain energies to less than 0.5 Ha/atom and pressures to 1.5\% error or less. Smaller discretization errors can be obtained as needed by increasing grid resolution, localization radius, and quadrature orders.

Whenever computationally feasible, we compare our SQDFT results with similar planewave results computed with VASP~\cite{Kresse1993a,Kresse1994,Kresse1996} and PWscf contained in the Quantum Espresso package~\cite{Giannozzi2009,Giannozzi2017}. The simulation parameters were chosen to allow a direct comparison to the SQDFT calculations and hence, the same convergence criteria and accuracy levels were applied. Both planewave codes were run between 2000~K and 200000~K with 64 atoms, the Baldereschi Mean value point, and the LDA exchange-correlation functional~\cite{Perdew1981}. We used the hard PAW pseudopotential in VASP, while we employed the same ONCV pseudopotential for the SQDFT and PWscf calculations. In the VASP calculations, we used a cutoff of 1000~eV and the PWscf simulations were run with 100~Ry and 400~Ry cutoffs for the wavefunctions and charge densities respectively. All planewave Kohn-Sham DFT-MD simulations were run for at least 20000 timesteps with timestep sizes between 0.1~fs and 0.4~fs depending on temperature. We applied a Nos\'{e}-Hoover thermostat in VASP, while PWscf was run with a Berendsen thermostat.

It is worth noting that though the SQ method is capable of performing simulations at low as well as high temperature \cite{Pratapa2016,Suryanarayana2018,Sharma2020}, we have restricted its usage to temperatures above 100~kK here. This is because the computational prefactor of the SQ method grows rapidly with decrease in temperature, making standard diagonalization-based methods/codes the more efficient choice for that regime. In particular, the required quadrature order has an inverse dependence on the temperature \cite{Suryanarayana2013} and the truncation radius also increases with decreasing temperature \cite{Suryanarayana2017}, i.e., the electronic interactions become more delocalized. It is also worth noting that the computational cost of the SQ method is not directly influenced by the density.

\section{Thermodynamic properties}
\subsection{Equation of state (EOS)}
%

The equation of state (EOS) data are directly calculated by averaging the thermodynamic properties pressure, energy, and temperature over the entire simulation length after a short equilibration period. 
In the following, we focus on the thermal equation of state and the Hugoniot, which were computed entirely within the Kohn-Sham framework. 

\begin{figure}[b]
  \includegraphics[width=1.0\linewidth]{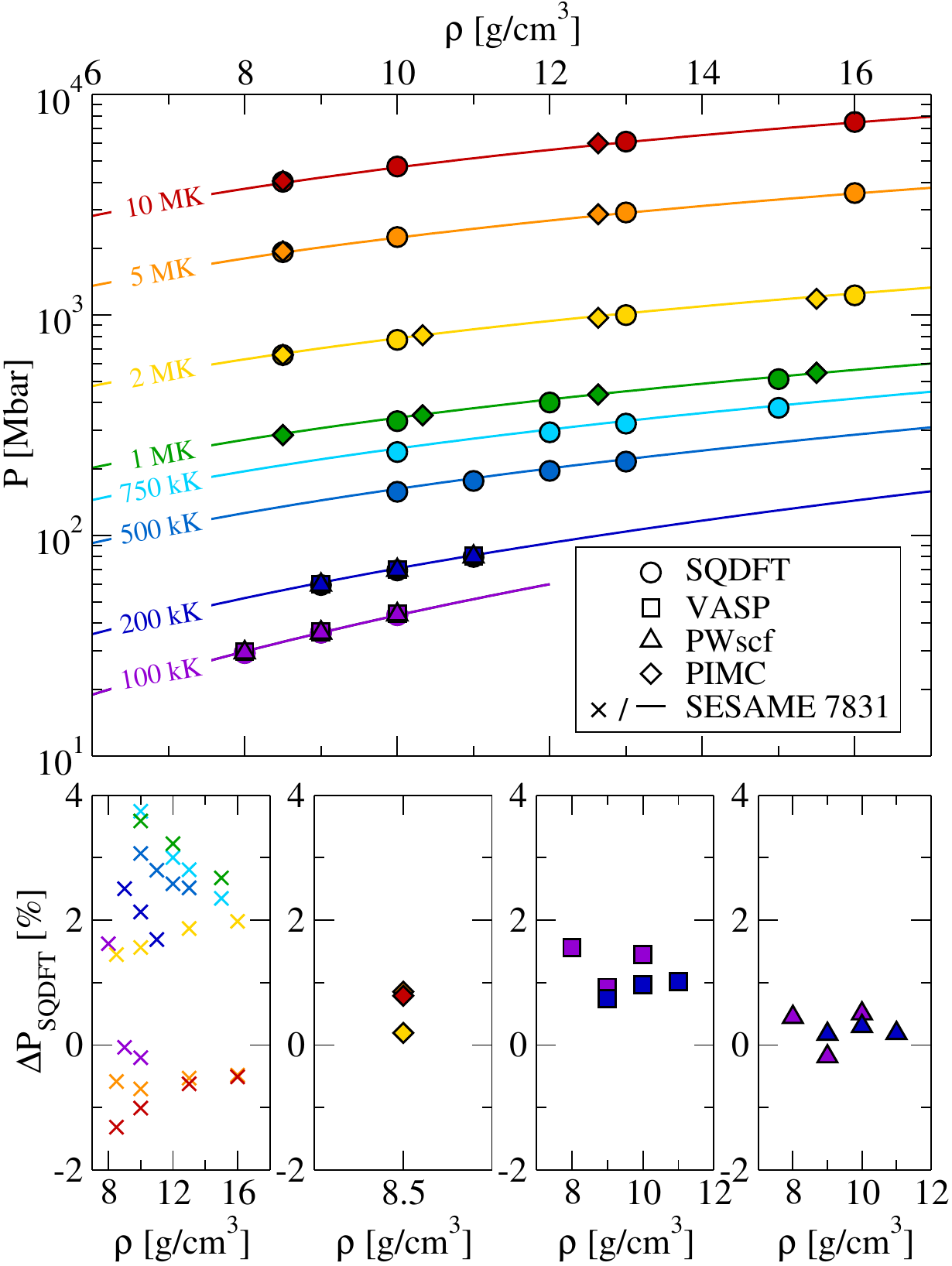}
  \caption{Comparison of the thermal equation of state for carbon based on different methods. Upper panel: Pressure calculated with SQDFT (circles) and the planewave Kohn-Sham DFT-MD codes PWscf (triangles) and VASP (squares). PIMC data (diamonds)~\cite{Benedict2014, Militzer2021} and the SESAME 7831 (colored lines)~\cite{Lyon1992} data are shown for comparison. Lower panels: Pressure difference between SQDFT and other approaches using the same color code and symbols as in the upper panel. 
}
  \label{fig1-EOS}
\end{figure} 

In Figure~\ref{fig1-EOS}, we present our results for the thermal equation of state that was calculated from 100~kK up to 10~MK and from 8~g/cm$^3$ to 16~g/cm$^3$ using the SQ method, as implemented in the SQDFT code. The plot also contains our benchmarking results using the planewave Kohn-Sham DFT-MD codes VASP and PWscf, as well as the wide-range EOS SESAME 7831 for liquid carbon~\cite{Lyon1992} and the PIMC data available in literature~\cite{Driver2012, Benedict2014, Militzer2021}. The upper panel illustrates the absolute pressures and the lower panels show the difference of SESAME 7831, PIMC, VASP, and PWscf relative to SQDFT from left to right. The color code in both panels refers to the temperatures indicated as numbers in the upper panel. Note that the SESAME results are shown as lines in the upper panel and as crosses in the lower left panel. 

The isotherms of the SESAME 7831 EOS model 
show overall a monotonic increase of pressure with increasing density. This behavior is recovered by all considered electronic structure codes, which are generally in good agreement as depicted by the maximum deviation of 4\% in the lower panels. These panels can be divided into two groups, i.e., on the left we present the comparison to two commonly used EOS for carbon at high temperatures and on the right we show the benchmark against widely used planewave Kohn-Sham DFT-MD codes using different types of pseudopotentials. The two rightmost panels show that the SQ method agrees well with the planewave results for 100~kK and 200~kK. The smallest deviations are found for PWscf, with EOS results differing by less than 
0.6\% from the SQDFT EOS results, while using the same ONCV pseudopotential as SQDFT. Note that the deviation could be further reduced 
by choosing stricter convergence criteria for both codes~\cite{Pratapa2016}. 
Comparing to the VASP code, which uses PAW rather than ONCV pseudopotentials, we find a slightly larger deviation of typically about 1\% and with a maximum of less than 2\%. Additionally, all values are positive, which indicates that the ONCV pseudopotential gives systematically slightly smaller values than the PAW pseudopotential. 
We find a significantly larger, yet still satisfactory, deviation of up to 4~\% comparing our SQDFT EOS results to the SESAME 7831 EOS model which is
constructed using a decomposition consisting of a temperature independent part (sometimes referred to as the cold curve), a ion thermal part and an electron thermal part based on average-atom (Inferno) DFT calculations~\cite{Liberman1982}. A reparametrization of that underlying model may yield closer agreement with full Kohn-Sham DFT.

For the systems where it can be performed, PIMC is thought to be the most accurate first-principles simulation technique to study the equilibrium properties of quantum systems in high temperature plasma states. It includes the effect of bonding, ionization, exchange-correlation and quantum degeneracy~\cite{Benedict2014}. However, this method becomes prohibitively expensive for systems with high atomic number and at lower temperatures when the free-particle nodal surface is not as good of an approximation and the sampling efficiency goes down.
With the SQ method we can reach temperatures where both Kohn-Sham DFT-MD and PIMC overlap. 
We find the Kohn-Sham DFT-MD pressures to be in excellent agreement with the PIMC data, with differences of less than 0.9\%, well within the targeted 1.5\% discretization error of the present calculations.
This comparison is crucial because the approximations inherent in Kohn-Sham DFT-MD and PIMC calculations are altogether different,
allowing us to verify that certain features of our results such as the location of the maximum compression in the Hugoniot curve (see Fig.~\ref{fig2-hugo})
are good indicators of the true EOS of carbon in the WDM regime.


\subsection{Hugoniot}

\begin{figure}[htb]
  \includegraphics[width=1.0\linewidth]{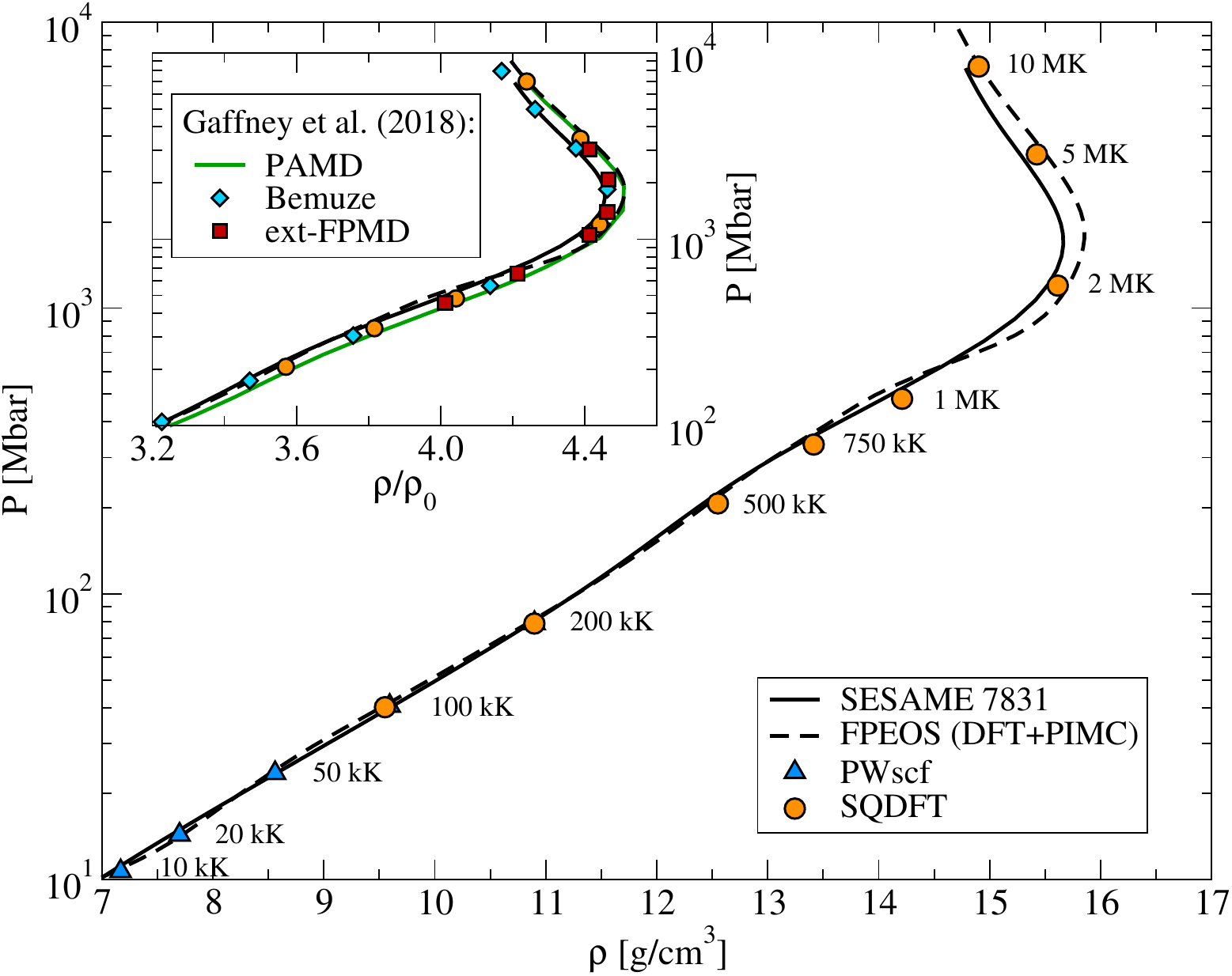}
  \caption{Hugoniot curve entirely based on Kohn-Sham DFT-MD by combining SQDFT (circles) and PWscf (triangles) as calculated in this work, compared to wide-range equations of state results using SESAME 7831 (solid line)~\cite{Lyon1992} and FPEOS (dashed line)~\cite{Militzer2021}. The indicated temperatures refer to the SQDFT and PWscf points. Inset: Comparison of pressure over compression ratio including pseudoatom models BEMUZE and PAMD as well as the extended FPMD method~\cite{Gaffney2018}.}
  \label{fig2-hugo}
\end{figure}

The equation of state data allow us to directly determine the Hugoniot curve by relating internal energy $u$, pressure $P$, and density $\rho$  of the initial (subscript 0) and shocked (subscript 1) state: 
\begin{equation}
   u_1 - u_0 = \frac{1}{2}(P_1 + P_0) (\frac{1}{\rho_0} - \frac{1}{\rho_1}) \text{.}
   \label{eq-hugo}
\end{equation}
The above equation is solved for every isotherm considering diamond at a density of 3.515~g/cm$^3$, a temperature of 300~K, and a pressure of 1 bar as an initial state. The internal energy of the initial state was determined to be -1249.378~kJ/g using PWscf with the 2s2p valence ONCV pseudopotential. The resulting full Kohn-Sham DFT-MD Hugoniot is plotted in Figure~\ref{fig2-hugo}. 
The blue triangles illustrate the planewave DFT-MD points between 10~kK and 200~kK and the orange circles represent the SQDFT results between 100~kK and 10~MK. Both codes are in very good agreement for the 200~kK point, while the PWscf point is slightly shifted towards a 0.04~g/cm$^3$ higher density at 100~kK, which results in a pressure deviation of 1\%. Overall, our results recover the general behavior of the  SESAME 7831 and FPEOS~\cite{Militzer2021} Hugoniots shown as solid and dashed lines, respectively, in Figure~\ref{fig2-hugo}. Note that the FPEOS combines planewave Kohn-Sham DFT-MD data at low \Rev{temperatures} with the PIMC data~\cite{Driver2012, Benedict2014} previously discussed in Figure~\ref{fig1-EOS} at high pressures. The SESAME and the PIMC Hugoniots are very similar up to 300~Mbar, but increasingly deviate for higher pressures with a maximum compression of 4.46 and 4.51, respectively. This compression maximum and subsequent slope change of the Hugoniot curve are due to the ionization of the 1s state, which is captured by all plotted data sets. Our Hugoniot curve agrees closely with the SESAME and FPEOS curves up to 200~kK and yields slightly lower pressures up to 750~kK. For the highest temperatures, our results fall between the FPEOS and SESAME curves. To further investigate this difference at high pressures, we compare our Hugoniot curve to other theories that are closely related to Kohn-Sham DFT in the inset of Figure~\ref{fig2-hugo}. In particular, we compare to the datasets gathered by Gaffney~\textit{et al.}~\cite{Gaffney2018}, i.e., extended FPMD and pseudoatom codes such as PAMD and BEMUZE. All approaches including ours reproduce the general trend induced by the 1s pressure ionization in this high-compression region and hence we cannot differentiate between those equations of state.
Considering the small difference observed in the compression maximum of these curves, it would be also extremely challenging to differentiate experimentally between the different approaches.

Nevertheless,  the excellent agreement of SQ with the PIMC data suggests that Kohn-Sham DFT-MD using the SQ method provides a promising approach to calculate the EOS of materials in the warm dense matter regime, in particular for higher atomic number systems out of reach for PIMC. 

\section{Structural properties}

\begin{figure}[h]
  \includegraphics[width=1.0\linewidth]{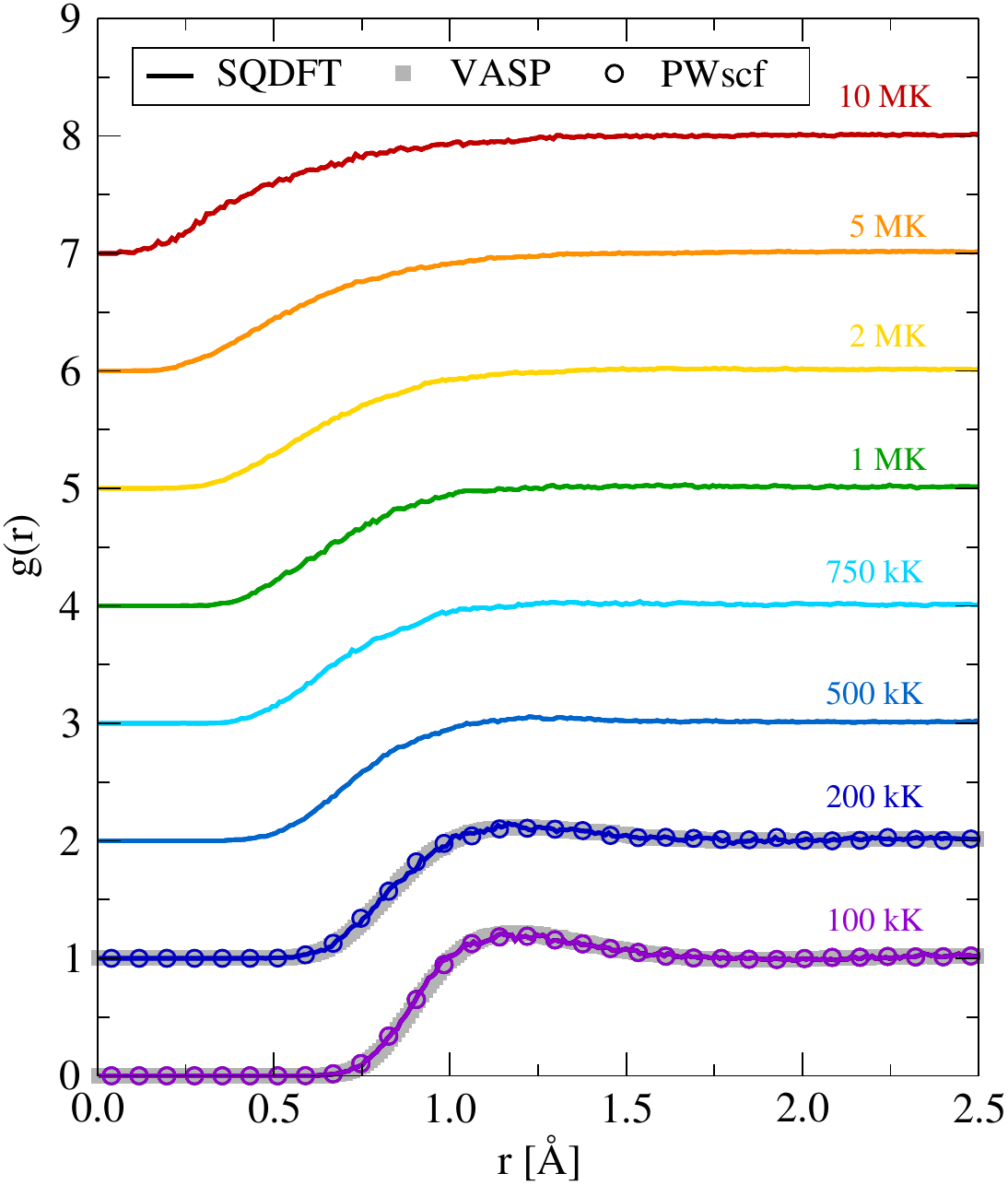}
  \caption{Pair distribution function of carbon obtained with SQDFT (colored lines) along the 10 g/cm$^3$ isochore between 100~kK and 10~MK. For comparison, the pair distribution functions calculated with VASP (filled gray squares) \Rev{and PWscf (open colored circles) are} shown for the two lowest temperatures. Note that the curves above 100~kK are shifted along the y-axis to improve visibility.}
  \label{fig3-pv}
\end{figure}

The structural properties of the high-temperature carbon plasma are accessible via the pair distribution function,

\begin{equation}
\Rev{
   g(r) = \frac{V}{4\pi r^2 N(N-1)}\bra \sum_{i=1}^N \sum_{\substack{j=1 \\ j \neq i}}^N \delta(r - |\vec{r}_i-\vec{r}_j|) \ket \text{, }}
   \label{eq-gr}
\end{equation}
\Rev{with particle number $N$, volume $V$, radial distance $r$ and the particle positions $\vec{r}_{i}$ and $\vec{r}_{j}$. The brackets $\bra \cdot \ket$ denote the time average and $\delta$ decribes the Dirac delta function. The results obtained from the molecular dynamics trajectories computed with SQ} along the 10 g/cm$^3$ isochore for all considered temperatures are shown in Figure~\ref{fig3-pv} as colored lines.

The 100~kK curve shows a broad peak at about 1.15~\AA, which indicates that the carbon atoms are still significantly correlated at this temperature. This broad peak is a remnant of a nearest-neighbor peak as it can be found in solids or liquids. The bonding distance of carbon is 1.54~\AA\, in uncompressed diamond, which shrinks to about 1.10~\AA\, at 10~g/cm$^3$ and 5000~K.

 The broad peak becomes less pronounced at 200~kK and shifts slightly towards smaller interatomic distances. As the temperatures increases further, the peak vanishes and the pair distribution functions extend to increasingly small distances. This behavior is typical for a rather weakly coupled plasma that is dominated by the kinetic energy and binary collisions, as is to be expected under these thermodynamic conditions with a compression factor of 2.85.

Additionally, the pair distribution functions calculated with VASP \Rev{and PWscf at 100~kK and 200~kK are shown} in Figure~\ref{fig3-pv} to benchmark our SQ results. The curves calculated with the \Rev{three} different codes agree very well\Rev{, as expected among systematically convergent methods. Particularly, the pair correlation functions computed with PWscf are almost indistinguishable from the respective SQ curves as these calculations employ the same ONCV pseudopotential.}
Note that the \Rev{planewave} simulations with 64 carbon atoms become too computationally demanding at 10~g/cm$^3$ above 200~kK and therefore, we rely entirely on SQ at high temperatures.

\begin{figure}[b]
  \includegraphics[width=1.0\linewidth]{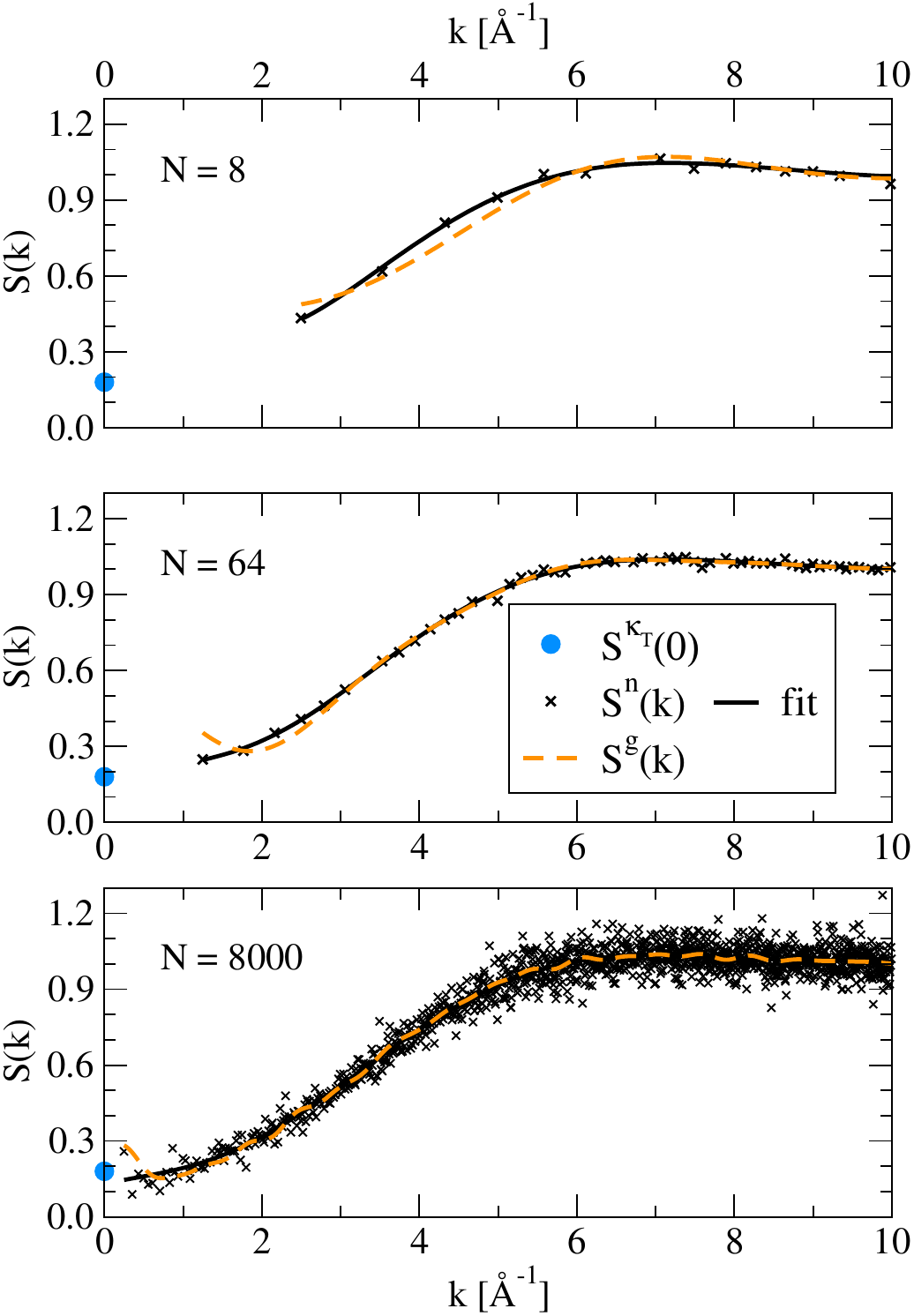}
  \caption{\Rev{Static structure factor for 8 (upper panel), 64 (middle panel), and 8000 (lower panel) carbon atoms at 10 g/cm$^3$ and 500~kK. The filled blue circles indicate the k=0 limit directly obtained from the equation of state results of the 64 atom calculations.}
}
  \label{fig4-strufak}
\end{figure} 

\Rev{The pair distribution function is not directly accessible in experiments, however, the closely related structure factor can be measured. In particular, the static ion-ion structure factor can be directly calculated by Fourier transforming the pair distribution function~\cite{Vorberger2020}:}
\begin{equation}
\Rev{
   S^{g}(k) = 1 + \frac{N}{V} \int_{-\infty}^{\infty} \mathrm{d}\vec{r}\, g(r) e^{\mathrm{i}\vec{k}\cdot \vec{r}}  \text{.}
}
   \label{eq-Sk-pdf}
\end{equation}
\Rev{The static ion-ion structure factor $S(k)$ can only be calculated for small wave vectors $k$, when the simulation box, and in turn the particle number, are chosen sufficiently large. Therefore, we exploit the capability of large-scale Kohn-Sham DFT-MD simulations performed by SQ and calculate the static ion-ion structure factor with up to 8000 carbon atoms. In Fig.~\ref{fig4-strufak}, we show the results for different particle numbers as dashed orange lines.} The 8000 atom calculation agrees well with the predictions based on smaller cell sizes, but one significant advantage becomes evident:  we can reach very small $k$ values that extrapolate nicely to the compressibility limit $S^{\kappa_T}(k=0) = \kappa_T N k_B T /V$, where $k_B$ is the Boltzmann constant and $\kappa_T$ is the compressibility. This is of crucial importance for X-ray Thomson scattering experiments~\cite{Ma2013, Saunders2018, Frydrych2020} and the measurement of ion acoustic modes~\cite{Rueter2014}. \Rev{The small $k$ behavior for a given simulation size can even be further improved by using an alternative method to calculate the static ion-ion structure factor}
\begin{equation}
\Rev{
   S^{n}(k) = \frac{1}{2\pi N} \int_{-\infty}^{\infty} \mathrm{d}\omega \int_{-\infty}^{\infty} \mathrm{d}t\, \bra n_{\vec{k}}(0)\, n_{-\vec{k}}(t) \ket e^{\mathrm{i}\omega t} \text{,}
}
   \label{eq-Sk-ion-density}
\end{equation}
\Rev{based on the correlation function of the Fourier-transformed ion density $n_{\vec{k}}(t) = \sum_{i = 1}^N e^{-\mathrm{i}\vec{k}\cdot \vec{r}_i(t)}$~\cite{Rueter2014}. The results and fits are shown as black crosses and solid lines in Fig.~\ref{fig4-strufak}.}

\section{Transport properties}
\subsection{Diffusion coefficient}
The diffusion coefficient is evaluated for each density-temperature point by integrating the velocity autocorrelation function:
\begin{equation}
   D = \frac{1}{3N} \int_0^{\infty} \mathrm{d}t\, \sum_{i=1}^N
   \bra \vec{v}_i(0) \cdot \vec{v}_i(t)\ket \text{,}
   \label{eq-diff}
\end{equation}
where $\vec{v}_i$ is the 3-dimensional velocity vector of the $i$-th particle.
\begin{figure}[h]
  \includegraphics[width=0.99\linewidth]{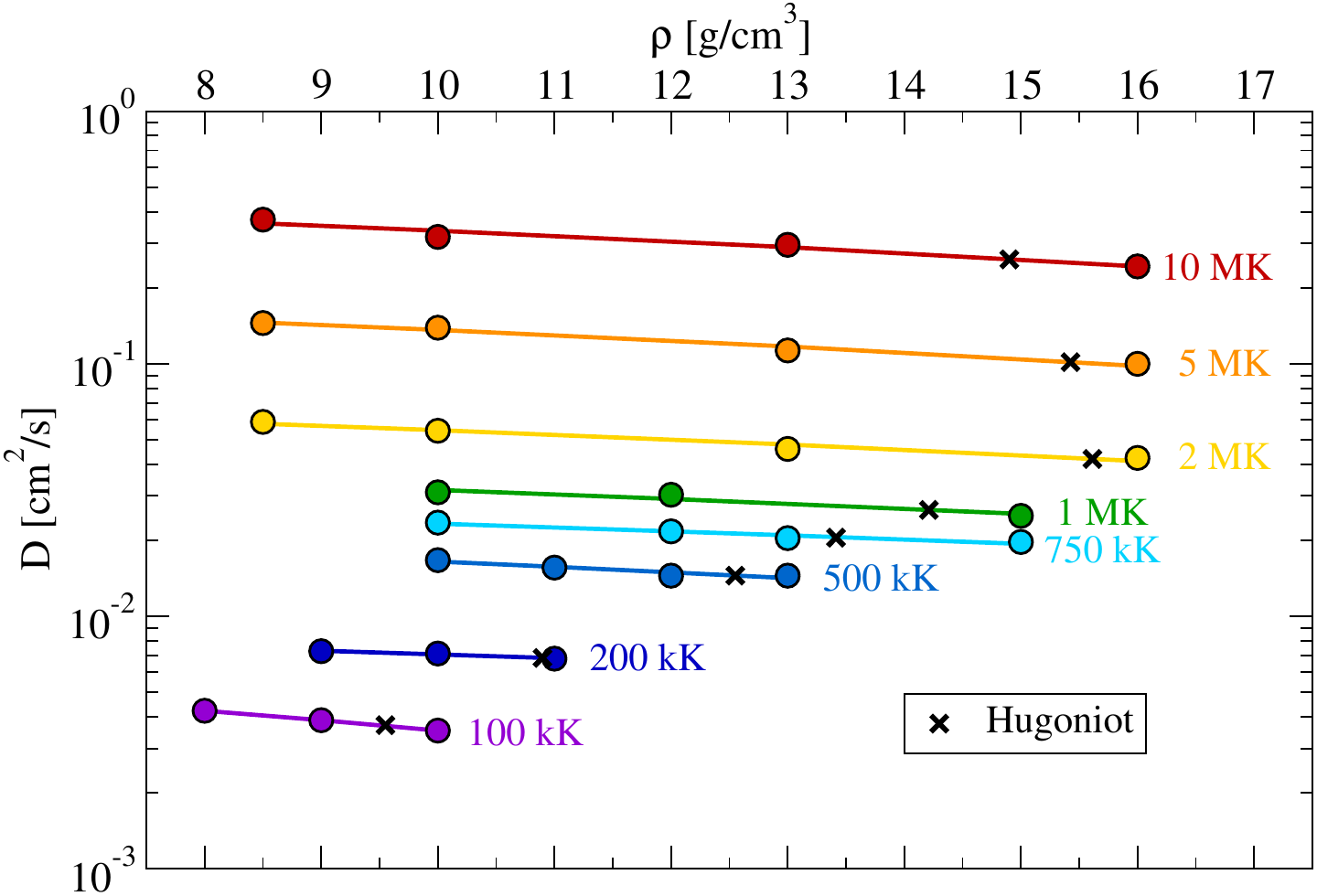}
  \includegraphics[width=1.00\linewidth]{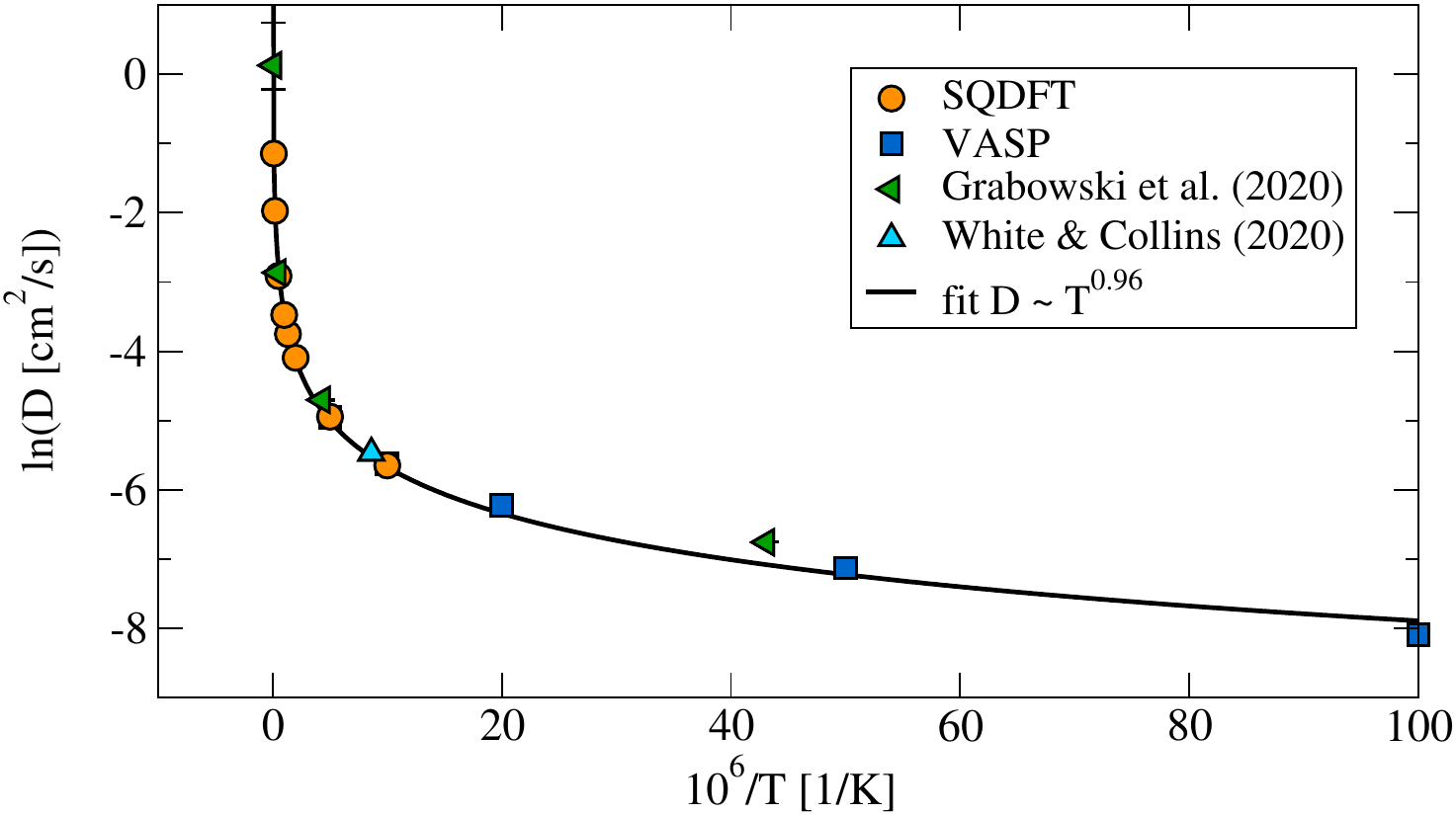}
  \caption{Diffusion coefficients calculated for the considered densities and temperatures. Top panel: Density dependence of the diffusion coefficients obtained with SQDFT for various temperatures up to 10~MK. The Hugoniot points are marked by black crosses. Bottom panel: \Rev{Arrhenius plot to illustrate the temperature behavior of} the diffusion coefficient along the 10~g/cm$^3$ isochore combining planewave and Spectral Quadrature DFT. For comparison, we show the result by White and Collins~\cite{White2020} as well as the dataset by Grabowski \textit{et al.}~\cite{Grabowski2020}. The solid black line illustrates the best T$^\alpha$ fit of our Kohn-Sham DFT-MD data.}
  \label{fig5-diff}
\end{figure}

The calculated diffusion coefficients are shown in Fig.~\ref{fig5-diff}. In the upper panel, we show the diffusion coefficients derived from the SQDFT trajectories as colored circles for all considered isotherms. The values along each isotherm decrease only slightly with density, which can be assumed to be linear over the considered density range. The linear fits of the diffusion coefficients are illustrated as colored curves. From these fits, we obtain the diffusion coefficients along the Hugoniot curve, which are indicated by black crosses.

In the lower panel, we show an Arrhenius plot of the diffusion coefficients obtained with SQDFT and VASP along the 10~g/cm$^3$ isochore. \Rev{We find our results to behave fundamentally differently compared to the Arrhenius law $\mathrm{ln}(D) \sim E_A/(k_B T)$ that breaks down over large temperature scales and in high-temperature systems with thermal energies that are significantly higher than the activation energy $E_A$. Therefore, we study} the temperature dependence of the \Rev{carbon diffusion coefficients along the 10~g/cm$^3$ isochore with a fit of the form $\mathrm{ln}(D) \sim -\alpha \mathrm{ln}(1/T)$, which is equivalent to $D \sim T^{\alpha}$ that has been previously used in high-temperature studies~\cite{Ticknor2015}. This functional form}
gives the known limiting cases of the Einstein-Stokes fluid for $\alpha = 1.0$ and the Maxwell-Boltzmann gas for $\alpha = 0.5$. We find a best fit value of \Rev{$\alpha=0.96$ considering the SQDFT and VASP results combined. The value is similar to the reported value of $\alpha=0.95$ for the heavy particles in hot dense HCNO plasmas considering temperatures up to 200~eV (2.32~MK)~\cite{Ticknor2015}}. Hence, the Kohn-Sham DFT-MD data agree with the expected Einstein-Stokes trend indicating a fluid-like behavior of the carbon plasma despite the high temperatures.
\Rev{Our} presented Kohn-Sham DFT-MD results in the lower panel of Fig.~\ref{fig5-diff} follow the trend of the data \Rev{discussed} in a comparative study by Grabowski \textit{et al.}\cite{Grabowski2020}, where the error bars indicate the spread of the predictions by the considered models and codes, which includes for example orbital-free MD and average atom models. Unfortunately, the datasets of the different approaches presented in that study are not available individually, so we cannot differentiate further. We can only compare directly to the mixed deterministic-stochastic DFT value at 10~g/cm$^3$ and 10~eV (116~kK) provided by White and Collins~\cite{White2020}, which is consistent with our calculations.
\Rev{Additionally, we find the diffusion coefficients} calculated with \Rev{the two different} Kohn-Sham DFT-MD codes at 100~kK and 200~kK agree within 5\%. \Rev{This agreement is at the level expected given the statistical errors in the SQ DFT-MD results (Tab.~\ref{tab:eos}), comparable errors in the VASP DFT-MD results, and different pseudopotentials employed in the SQ and VASP calculations.}

\subsection{Viscosity}
 The viscosity $\eta$ is calculated by integrating the ensemble average of the autocorrelation functions defined via the stress tensor:
\begin{equation}
   \eta = \frac{V}{5k_B T} \int_0^{\infty} \mathrm{d}t \sum_{i=1}^5 \bra \sigma_{i}(0)\cdot \sigma_{i}(t)\ket \text{.}
   \label{eq-eta}
\end{equation}
The five individual autocorrelation functions are given by the three off-diagonal stress tensor components $\sigma_{xy}$, $\sigma_{yz}$, $\sigma_{zx}$, and the linear combination of the diagonal components $(\sigma_{xx}-\sigma_{yy})/2$ and $(\sigma_{yy}-\sigma_{zz})/2$~\cite{Alfe1998}.
\begin{figure}[b]
  \includegraphics[width=1.0\linewidth]{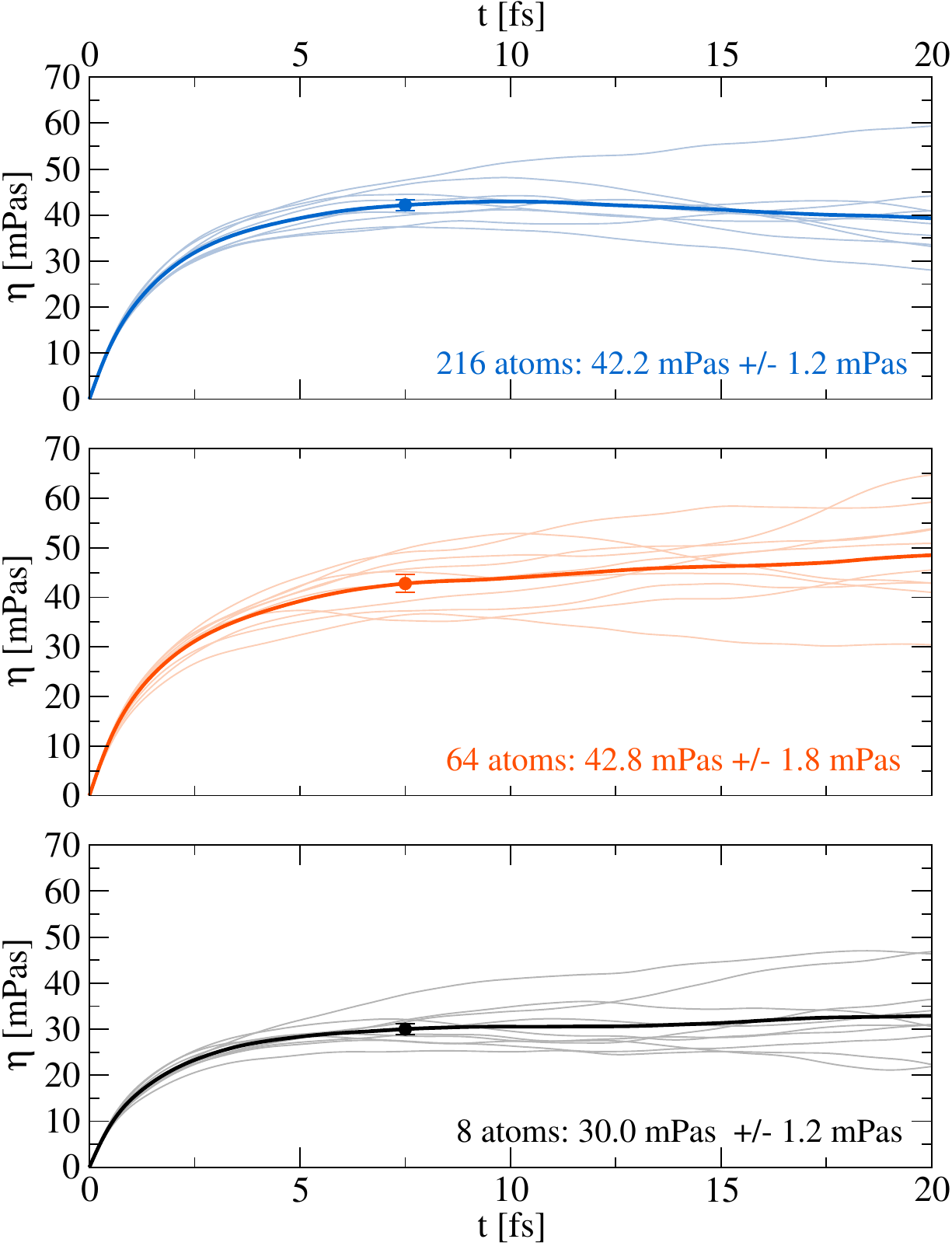}
  \caption{Viscosity of carbon at 16~g/cm$^3$ and 2~MK calculated for different unit cell sizes containing 216 (top), 64 (middle), and 8 (bottom) atoms. The thin lines represent the individual SQDFT simulations, whose average is shown as thick line. The final values for each case are given as filled circles \Rev{with $1\sigma$ error bars}.}
  \label{fig6-eta}
\end{figure}
Viscosity converges in general much more slowly than the diffusion coefficient and hence, we only demonstrate its calculation and the important finite-size effect for one plasma condition. In particular, we evaluate the viscosity at 16~g/cm$^3$ and 2~MK, which is close to the maximum compression of our calculated Hugoniot curve (see Fig.~\ref{fig2-hugo}). We consider different system sizes containing 8, 64, and 216 carbon atoms. For each system size, we use 10 different starting configurations and run every individual simulation for 24000 timesteps. The results of those individual runs are shown as thin lines in Fig.~\ref{fig6-eta} and their averages are illustrated as thick lines. All thin curves show a very strong variation independent of the considered particle number. This shows the importance of sampling viscosity properly by running multiple simulations or one very long simulation with more than 100000 timesteps. 
The final viscosity value for each system is marked as filled circle at \Rev{$t=$}7.5~fs. The calculated values of \Rev{42.2~mPas} and 42.8~mPas for 216 and 64 atoms, respectively, agree very well within the error bars. At the same time, we find a value of 30.0~mPas for 8 atoms, which deviates more than 25\% from the converged values. Therefore, we conclude that 8 atoms are not sufficient to study viscosity at these high-temperature conditions, even though the plasma is only mildly correlated and almost classical.

\section{Conclusion}
%
In this work, we studied the thermodynamic and transport properties of carbon up to temperatures of 10 million kelvin using full Kohn-Sham density functional theory molecular dynamics. By employing the Spectral Quadrature method, we are able to cover the principal carbon Hugoniot spanning conditions from the non-classical to the classical plasma regime, all at the Kohn-Sham level of theory. Previous such ab initio studies employing conventional planewave based Kohn-Sham methods were restricted to  sufficiently high densities at temperatures above 1~MK or temperatures well below the Fermi temperature for typical densities along the Hugoniot.



In the considered thermodynamic range, we find our EOS results in very good agreement with planewave Kohn-Sham DFT-MD, reproducing PWscf pressures along the 100~kK and 200~kK isotherms within 0.6~\%. And we find excellent agreement with PIMC results at temperatures of 10~MK and above. This is particularly notable because the theoretical approach to solve the many-particle problem in PIMC is inherently different and both approaches rely on complementary approximations (DFT exchange-correlation approximation vs.~PIMC fixed node approximation). Yet, they agree to within 0.9\% for the thermal EOS leading to a similar description of the Hugoniot, and giving a measure of the uncertainty in the EOS of warm dense carbon. 

One of the major benefits of an efficient many-particle method is the ability to generate ionic structural and transport properties by performing molecular dynamics simulations, e.g., pair distribution function and diffusion coefficients. For carbon under the conditions studied here, these properties reflected the nature of a liquid-like and rather weakly coupled carbon plasma that reproduces the Einstein-Stokes diffusion behavior. Furthermore, the $O(N)$ scaling of the SQ method allows us to treat large particle numbers. Hence, we are able to explore properties that are notoriously hard to converge, such as viscosity, and access the low-$k$ limit of the ion-ion structure factor, which in turn is related to the isothermal compressibility. The structure factor can be measured in X-ray Thomson scattering experiments, which allow the derivation of plasma parameters such as temperature, density, and ionization state. Therefore, the approach used here may provide useful information in the interpretation of such measurements.

Some of the results presented in this work can likely be obtained by computationally less expensive approaches such as Average Atom models or the Hypernetted Chain Approximation \cite{DharmaWardana2008, Bredow2013}, which are expected to provide a faithful representation of the EOS for temperatures sufficiently above the Fermi temperature~\cite{Gaffney2018}. However, as with all more approximate methods, their accuracy and region of applicability are not known a priori. Fully ab initio calculations as presented here hence provide important benchmarks to clarify the accuracy and applicability of more approximate methods, as well as providing key data which may be used to inform and improve such methods.

Future work will be directed towards other ablator materials for inertial confinement fusion experiments such as beryllium and hydrocarbons and higher-Z materials such as iron and nickel. 
%
\begin{acknowledgments}
We gratefully acknowledge D. R. Hamann for his invaluable assistance in the construction of robust all-electron pseudopotentials. This work was performed under the auspices of the U.S. Department of Energy by Lawrence Livermore National Laboratory under Contract DE-AC52-07NA27344. M.B. was partially supported by the European Union within the Marie Sk{\l}odowska-Curie actions (xICE grant 894725). LDRD 16-ERD-011 provided partial support. All simulations were carried out on the supercomputer Quartz at LLNL. Computing support for this work came in part from the Lawrence Livermore National Laboratory (LLNL) Institutional Computing Grand Challenge program.
\end{acknowledgments}

%

\appendix
\section{SQDFT data}
The EOS data and diffusion coefficients computed for 64 carbon atoms with SQDFT are summarized in Tab.~\ref{tab:eos}. The data of the lowest three isotherms were obtained with the 4-electron ONCV pseudopotential, while the higher temperature data were calculated with the all-electron ONCV pseudopotential. The energies of the all-electron potential are shifted by a constant energy of 6885.81~kJ/g to match the Hugoniot starting condition, consistent with the 4-electron pseudopotential.

\begin{table*}
    \centering
    \begin{tabular}{c | c | c | c | c | c | c | c}
        $\rho$[g/cm$^3$] & T[K] & P[Mbar] & \Rev{P$_{\mathrm{err}}$[Mbar]} & u[kJ/g] & \Rev{u$_{\mathrm{err}}$[kJ/g]} & $D$[cm$^2$/s] & \Rev{$D_{\mathrm{err}}$[cm$^2$/s]}\\
        \hline
        8.0  & 100000 & 29.237 & 0.025 & -914.08 & 0.38 & 0.00423 & 0.00012 \\
        9.0  & 100000 & 36.183 & 0.020 & -897.09 & 0.25 & 0.00388 & 0.00012 \\
        10.0 & 100000 & 43.461 & 0.024 & -878.64 & 0.26 & 0.00353 & 0.00010 \\
        \hline
        9.0  & 200000 & 59.514 & 0.013 & -517.70 & 0.17 & 0.00728 & 0.00022\\
        10.0 & 200000 & 69.299 & 0.020 & -504.75 & 0.40 & 0.00713 & 0.00027\\
        11.0 & 200000 & 79.855 & 0.018 & -488.49 & 0.18 & 0.00680 & 0.00036\\
        \hline
        10.0 & 500000 & 157.68 & 0.02 & 853.47 & 0.29 & 0.0167& 0.0005\\
        11.0 & 500000 & 176.62 & 0.03 & 858.82 & 0.29 & 0.0156& 0.0005\\
        12.0 & 500000 & 196.24 & 0.03 & 866.98 & 0.32 & 0.0145& 0.0004\\
        13.0 & 500000 & 216.17 & 0.04 & 875.03 & 0.46 & 0.0145& 0.0006\\
        \hline
        10.0 & 750000 & 238.84 & 0.03 & 2280.0 & 0.4 & 0.0235& 0.0013\\
        12.0 & 750000 & 293.56 & 0.06 & 2256.7 & 0.7 & 0.0217& 0.0006\\
        13.0 & 750000 & 321.45 & 0.06 & 2248.1 & 0.7 & 0.0204& 0.0010\\
        15.0 & 750000 & 379.25 & 0.06 & 2245.0 & 0.6 & 0.0197& 0.0010\\
        \hline
        10.0 & 1000000 & 329.96 & 0.07 & 4034.6 & 0.9 & 0.0310& 0.0019\\
        12.0 & 1000000 & 400.99 & 0.08 & 3946.0 & 0.9 & 0.0303& 0.0014\\
        15.0 & 1000000 & 511.61 & 0.08 & 3866.5 & 0.8 & 0.0250& 0.0012\\
        \hline
        8.5  & 2000000 & 658.41 & 0.04 & 12985 & 0.7 & 0.0589 & 0.0003\\
        10.0 & 2000000 & 771.56 & 0.05 & 12681 & 0.8 & 0.0543 & 0.0006\\
        13.0 & 2000000 & 998.99 & 0.16 & 12215 & 1.8 & 0.0460 & 0.0009\\
        16.0 & 2000000 & 1228.3 & 0.3  & 11872 & 2 & 0.0424 & 0.0010\\
        \hline
        8.5  & 5000000 & 1928.9 & 0.2 & 37853 & 4 & 0.145& 0.004\\
        10.0 & 5000000 & 2259.9 & 0.2 & 37483 & 4 & 0.139& 0.006\\
        13.0 & 5000000 & 2920.3 & 0.3 & 36886 & 3 & 0.113& 0.003\\
        16.0 & 5000000 & 3578.2 & 0.3 & 36393 & 3 & 0.100& 0.003\\
        \hline
        8.5  & 10000000 & 4023.2 & 0.3 & 75558 & 5 & 0.372& 0.017\\
        10.0 & 10000000 & 4722.2 & 0.4 & 75222 & 7 & 0.318& 0.023\\
        13.0 & 10000000 & 6120.2 & 0.6 & 74705 & 8 & 0.296& 0.020\\
        16.0 & 10000000 & 7513.8 & 0.7 & 74236 & 8 & 0.243& 0.009\\
    \end{tabular}
    \caption{SQDFT equation of state data (density $\rho$, temperature $T$, pressure $P$, internal energy $u$) and diffusion coefficients $D$. \Rev{The 1$\sigma$ errors $P_{err}$, $u_{err}$, and $D_{\mathrm{err}}$ are given for the pressure, internal energy and diffusion coeffients, respectively.}}
    \label{tab:eos}
\end{table*}
%
\bibliography{sqdft}

\end{document}